\documentclass[11pt]{article}
\usepackage{moriond,epsfig}

\def\be{\begin{equation}}
\def\ee{\end{equation}}
\def\bea{\begin{eqnarray}}
\def\eea{\end{eqnarray}}

\begin{document}

\newcommand{\half}{{1\over2}}
\newcommand{\nad}{n_{\rm ad}}
\newcommand{\niso}{n_{\rm iso}}
\newcommand{\fiso}{f_{\rm iso}}
\newcommand{\ii}{\'{\'i}}
\newcommand{\bk}{{\bf k}}
\newcommand{\Ocdm}{\Omega_{\rm cdm}}
\newcommand{\ocdm}{\omega_{\rm cdm}}
\newcommand{\OM}{\Omega_{\rm M}}
\newcommand{\OB}{\Omega_{\rm B}}
\newcommand{\oB}{\omega_{\rm B}}
\newcommand{\OL}{\Omega_\Lambda}
\newcommand{\cltt}{C_l^{\rm TT}}
\newcommand{\clte}{C_l^{\rm TE}}
\newcommand{\calR}{{\cal R}}
\newcommand{\calS}{{\cal S}}
\newcommand{\Rrad}{{\cal R}_{\rm rad}}
\newcommand{\Srad}{{\cal S}_{\rm rad}}

\vspace*{4cm}
\title{BOUNDS ON ISOCURVATURE PERTURBATIONS FROM CMB AND LSS}

\author{ J. GARC\'IA-BELLIDO }

\address{Departmento de F\'{\i}sica Te\'orica, 
Universidad Aut\'onoma de Madrid\\
Cantoblanco, 28049 Madrid, Spain}

\maketitle\abstracts{ Using both cosmic microwave background and large
scale structure data, we put stringent bounds on the possible cold
dark matter, baryon and neutrino isocurvature contributions to
primordial fluctations in the Universe. Neglecting the possible
effects of spatial curvature, tensor perturbations and reionization,
we perform a Bayesian likelihood analysis with nine free parameters,
and find that the relative amplitude of the isocurvature component
cannot be larger than about a few \% for uncorrelated models. On the
other hand, for correlated adiabatic and isocurvature components, the
ratio could be slightly larger. However, the cross-correlation
coefficient is strongly constrained, and maximally
correlated/anticorrelated models are disfavored.}


Thanks to the tremendous developments in
observational cosmology during the last few years, it is possible to
speak today of a Standard Model of Cosmology, whose parameters are
known within systematic errors of just a few percent. Moreover, the
recent measurements of both temperature and polarization anisotropies
in the cosmic microwave background (CMB) has opened the possibility to
test not only the basic paradigm for the origin of structure, namely
inflation, but also the precise nature of the primordial fluctuations
that gave rise to the CMB anisotropies and the density perturbations
responsible for the large scale structure (LSS) of the Universe.

The simplest realizations of the inflationary paradigm predict an
approximately scale invariant spectrum of adiabatic and Gaussian
curvature fluctuations, whose amplitude remains constant outside the
horizon, and therefore allows cosmologists to probe the physics of
inflation through observations of the CMB anisotropies and the LSS
matter distribution. However, this is certainly not the only
possibility.  Multiple-field inflation predicts that, together with
the adiabatic component, there should also be entropy or
isocurvature
perturbations~\cite{Linde:1985yf,Polarski:1994rz,Garcia-Bellido:1995qq,Gordon:2000hv,Wands:2002bn,Finelli:2000ya},
associated with fluctuations in number density between different
components of the plasma before decoupling, with a possible
statistical correlation between the adiabatic and isocurvature
modes~\cite{Langlois:1999dw}.  Baryon and cold dark matter (CDM)
isocurvature perturbations were proposed long
ago~\cite{Efstathiou:1986} as an alternative to adiabatic
perturbations.  Recently, two other modes, neutrino isocurvature
density and velocity perturbations, have been added to the
list~\cite{Bucher:1999re}. Topological defects also predict a
significant isocurvature component. Moreover, it is well known that
entropy perturbations seed curvature perturbations outside the
horizon~\cite{Polarski:1994rz,Gordon:2000hv}, so it is possible that a
significant component of the observed adiabatic mode could be strongly
correlated with an isocurvature mode. Such models are generically
called {\em curvaton models}~\cite{Lyth:2001nq}, and are now widely
studied as an alternative to the standard paradigm. Furthermore,
isocurvature modes typically induce non-Gaussian signatures in the
spectrum of primordial perturbations.

I will review here our recent results~\cite{Crotty:2003aa}
constraining the various isocurvature components, using data from the
temperature power spectrum and temperature-polarization
cross-correlation recently measured by the WMAP satellite~\cite{WMAP};
from the small-scale temperature anisotropy probed by
ACBAR~\cite{ACBAR}; and from the matter power spectrum measured by the
2-degree-Field Galaxy Redshift Survey (2dFGRS)~\cite{2dFGRS}. We do
not use the data from Lyman-$\alpha$ forests, since they are based on
non-linear simulations carried under the assumption of adiabaticity.
We will not assume any specific model of inflation, or any particular
mechanism to generate the perturbations (late decays, phase
transitions, cosmic defects, etc.), and thus will allow all five modes
-- adiabatic (AD), baryon isocurvature (BI), CDM isocurvature (CDI),
neutrino isocurvature density (NID) and neutrino isocurvature velocity
(NIV) -- to be correlated (or not) among eachother, and to have
arbitrary tilts. However, we will only consider the mixing of the
adiabatic mode and one of the isocurvature modes at a time. This
choice has the advantage of restricting considerably the parameter
space, and reflects the fact that most of the proposed mechanisms for
the generation of isocurvature perturbations lead to only one
mode. The first bounds on isocurvature perturbations assumed
uncorrelated modes~\cite{Stompor:1995py}, but recently also correlated
ones were considered in
Refs.~\cite{Trotta:2001yw,Amendola:2001ni,Gordon:2002gv,Valiviita:2003ty}.

For simplicity, we also neglect the possible effects of spatial
curvature, tensor perturbations and reionization. Therefore, each
model is described by nine parameters: the cosmological constant
$\OL$, the baryon density $\oB=\OB h^2$, the cold dark matter density
$\ocdm=\Ocdm h^2$, the overall normalisation $A$, the relative
amplitude $\alpha$ of the isocurvature to adiabatic modes, and their
correlation $\beta$, the adiabatic and isocurvature tilts, $\nad$ and
$\niso$, and finally a matter bias $b$ associated to the 2dF power
spectrum. We generate a 5-dimensionnal grid of models ($A$, $\alpha$,
$\beta$, and $b$ are not discretized) and perform of Bayesian analysis
in the full 9-dimensional parameter space. At each grid point, we
store some $C_l$ values in the range $0 < l < 1800$ and some $P(k)$
values in the range probed by the 2dF data. The likelihood of each
model is then computed using the software or the detailed information
provided on the experimental websites, using 1398 points from WMAP, 11
points from ACBAR and 32 points from the 2dFGRS. For parameter values
which do not coincide with grid points, our code first performs a
cubic interpolation of each power spectrum, accurate to better than
one percent, and then computes the likelihood of the corresponding
model.  From the limitation of our grid, we impose a flat prior on the
isocurvature tilt: $\niso > 0.6$ -- inflation predicts the tilts of both
adiabatic and isocurvature modes to be close to one. The other grid
ranges are wide enough in order not to affect our results.

For the theoretical analysis, we will use the notation and some of the
approximations of Refs.~\cite{Gordon:2000hv,Wands:2002bn}. During
inflation more than one scalar field could evolve sufficiently slowly
that their quantum fluctuations perturb the metric on scales larger
than the Hubble scale during inflation. These perturbations will later
give rise to one adiabatic mode and several isocurvature modes. We
will restrict ourselves here to the situation where there are only two
fields, $\phi_1$ and $\phi_2$, and thus only one isocurvature and one
adiabatic mode. The evolution during inflation will draw a trajectory
in field space.  Perturbations along the trajectory (i.e. in the
number of e-folds N) will give rise to curvature perturbations $\calR_k
= \delta N_k = H\,\delta_k = H\,\delta\rho_k/\dot\rho$ on comoving
hypersurfaces, while perturbations orthogonal to the trajectory will
give rise to gauge invariant entropy (isocurvature) perturbations,
$\calS = \delta\ln(n_i/n_j) = \delta\rho_i/(\rho_i+p_i) -
\delta\rho_j/(\rho_j+p_j)$.\footnote{For instance, entropy perturbations in
cold dark matter during the radiation era can be computed as
$\calS_{\rm CDI} = \delta_m - 3\delta_r/4$.}

In order to relate these perturbations during inflation with those
produced during the radiation era, one has to follow the evolution
across reheating. During inflation we can always perform an
instantaneous rotation along the field trajectory and relate
perturbations in the fields, $\delta\phi_1$ and $\delta\phi_2$, with
perturbations along and orthogonal to the trajectory,
$\delta\sigma$ and $\delta s$,
\begin{equation}
\left(\begin{array}{c}\delta\sigma\\[2mm] \delta s\end{array}\right) =
\left(\begin{array}{lr}\cos\theta & \sin\theta\\[2mm] -\sin\theta & 
\cos\theta\end{array}\right)
\left(\begin{array}{c}\delta\phi_1\\[2mm] \delta\phi_2\end{array}\right)\,.
\end{equation}
In this case, the curvature and entropy perturbations can be written,
in the slow-roll approximation, 
as~\cite{Garcia-Bellido:1995qq,Gordon:2000hv}
\begin{eqnarray}
&&\calR_k = H\,{\dot\phi_1\delta\phi_1+\dot\phi_2\delta\phi_2\over
\dot\phi_1^2+\dot\phi_2^2} = H\,{\delta\sigma_k\over\dot\sigma}\,, \\[2mm]
&&\,\calS_k = H\,{\dot\phi_1\delta\phi_2-\dot\phi_2\delta\phi_1\over
\dot\phi_1^2+\dot\phi_2^2} = H\,{\delta s_k\over\dot\sigma}\,. 
\end{eqnarray}

The problem however is that, contrary to the case of purely adiabatic
perturbations, the amplitude of the curvature and entropy
perturbations does not remain contant on superhorizon scales, but
evolve with time.  In particular, due to the conservation on the
energy momentum tensor, the entropy perturbation seeds the curvature
perturbation, and thus their amplitude during the radiation dominated
era evolves according to
\begin{eqnarray}\label{evolution}
&&\dot\calR_k = \alpha(t)H\,\calS_k\,, \\[2mm]
&&\,\dot\calS_k = \beta(t)H\,\calS_k\,,
\end{eqnarray}
where $\alpha$ and $\beta$ are time-dependent functions characterizing 
the evolution during inflation and radiation eras. A formal solution
can be found in terms of a transfer matrix, relating the amplitude at
horizon crossing during inflation (*) with that at a later time during
radiation,
\begin{equation}
\left(\begin{array}{c} \calR \\[2mm] \calS \end{array}\right) =
\left(\begin{array}{lr} 1 & T_{\calR\calS}\\[2mm]
 0 & T_{\calS\calS}\end{array}\right)
\left(\begin{array}{c} \calR_* \\[2mm] \calS_* \end{array}\right)\,,
\end{equation}
where the transfer functions are given by
\begin{eqnarray}\label{TF}
&&T_{\calS\calS}(t,t_*) = 
\exp\left[\int_{t_*}^t\beta(t')H(t')dt'\right]\,, \\[2mm]
&&T_{\calR\calS}(t,t_*) = 
\int_{t_*}^t\alpha(t')H(t')T_{\calS\calS}(t',t_*)dt'\,.
\end{eqnarray}
Note that in the absence of primordial isocurvature perturbation,
$\calS_*=0$, the curvature perturbation remains constant and no
isocurvature perturbation is generated during the evolution. This
is the reason for the entries $T_{\calR\calR}=1$ and 
$T_{\calS\calR}=0$, respectively, in the transfer matrix.

Since $\phi_1$ and $\phi_2$ are essentially massless during inflation,
we can treat them as free fields whose fluctuations at horizon
crossing have an amplitude $\delta\phi_i =
(H_k/\sqrt{2k^3})\,e_i(\bk)$, where $H_k$ is the rate of expansion at
the time the perturbation crossed the horizon $(k_*=aH)$, and
$e_i(\bk)$ are gaussian random fields with zero mean, $\langle
e_i(\bk)\rangle=0$ and $\langle e_i(\bk)e_j^*(\bk')\rangle
=\delta_{ij}\,\delta(\bk-\bk')$. Now, since $\delta\sigma_k$ and
$\delta s_k$ and just rotations of the field fluctuations, they are
also gaussian random fields of amplitude $H_k/\sqrt{2k^3}$. However,
the time evolution (\ref{evolution}) will mix those modes and will
generically induce correlations and non-gaussianities.

Therefore, the two-point correlation function or power spectra of both
adiabatic and isocurvature perturbations, as well as their
cross-correlation, can be parametrized with three power laws, i.e.
three amplitudes and two spectral indices,
\begin{eqnarray}\nonumber
\Delta_\calR^2(k)\!&\equiv&\!
{k^3\over2\pi^2}\langle\calR^2\rangle = 
A^2\,\left({k\over k_0}\right)^{\nad-1}\,,\\[2mm]
\Delta_\calS^2(k)\!&\equiv&\!
{k^3\over2\pi^2}\langle\calS^2\rangle = 
B^2\,\left({k\over k_0}\right)^{\niso-1}\,,\\[2mm]
\Delta_{\calR\calS}^2(k)\!&\equiv&\!{k^3\over2\pi^2}\nonumber
\langle\calR\calS\rangle = A\,B\,\cos\Delta\ 
\left({k\over k_0}\right)^{(\nad+\niso)/2-1}\,,
\end{eqnarray}
where $k_0$ is some pivot scale, typically of the order of the present
horizon. Note that we have assumed here that the correlation coefficient
$\cos\Delta$ is scale-independent. In general there will be a small
scale dependence coming from that of $t_*(k)$ in the transfer
functions (\ref{TF}).  We will ignore it for the moment.\footnote{In a
recent paper~\cite{Valiviita:2003ty} this assumption was relaxed, but
produced results that are not very sensitive to it.}

\begin{figure}[t]
\psfig{figure=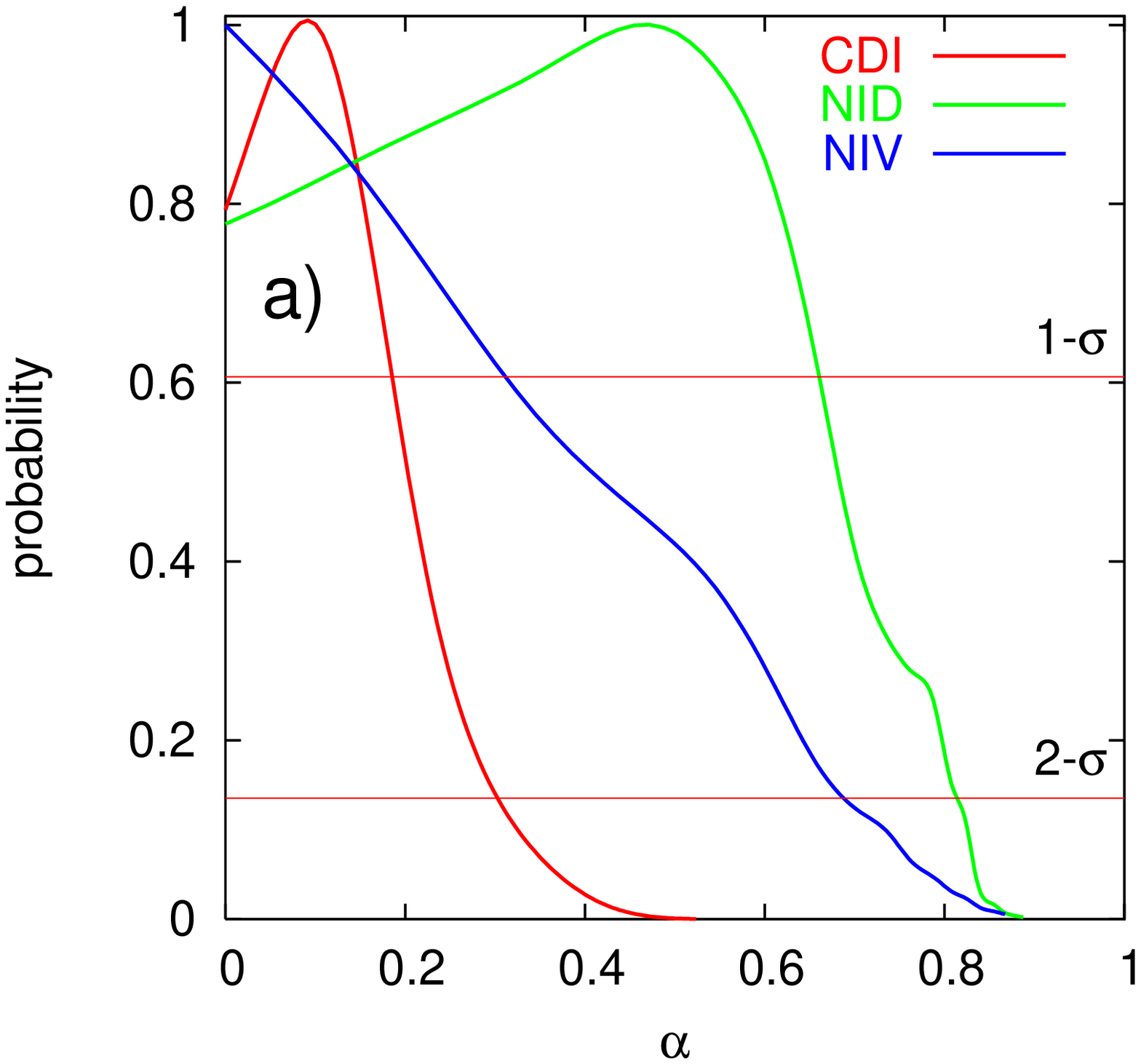,angle=-90,width=7cm}
\psfig{figure=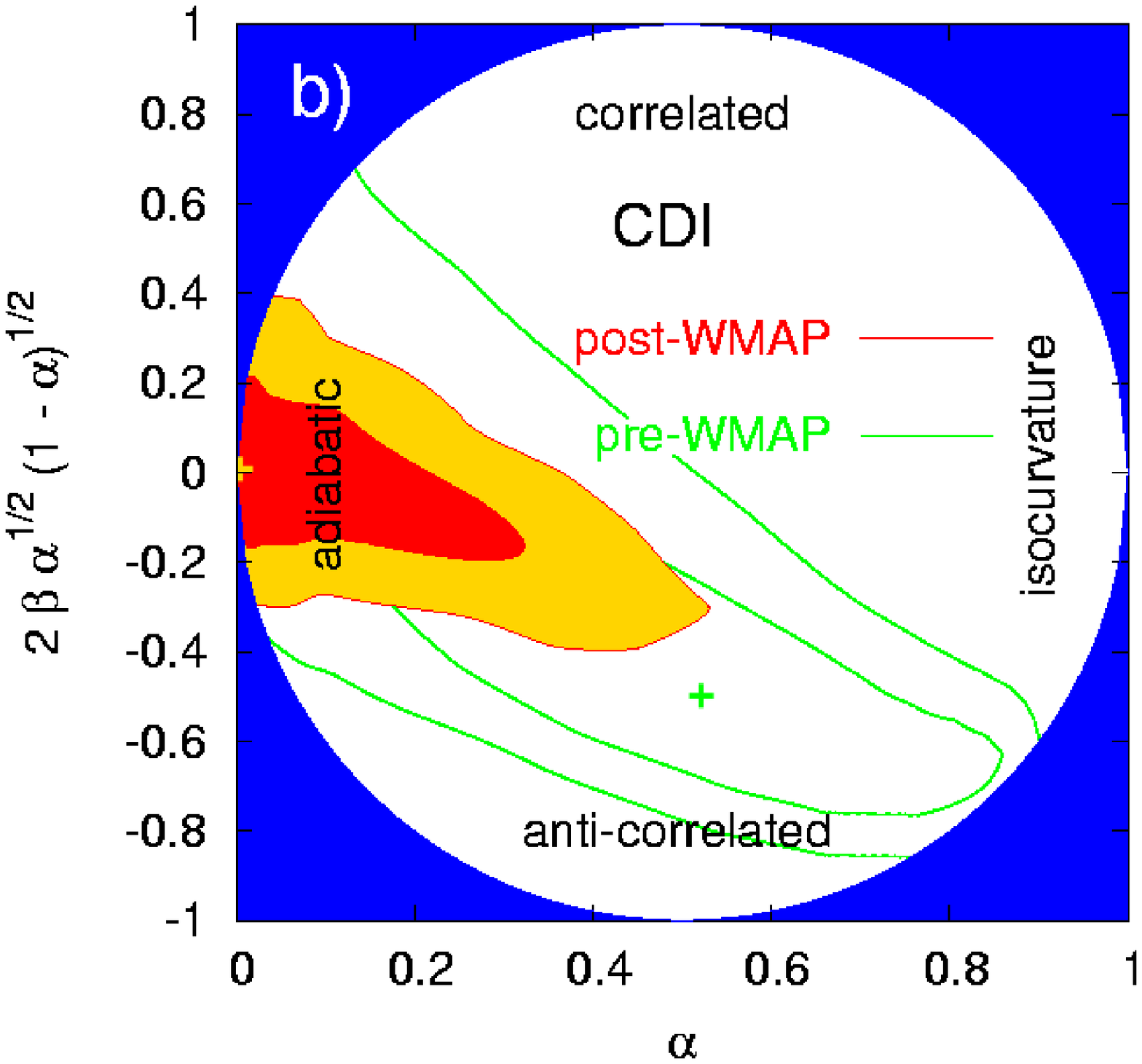,angle=-90,width=7cm}\\
\psfig{figure=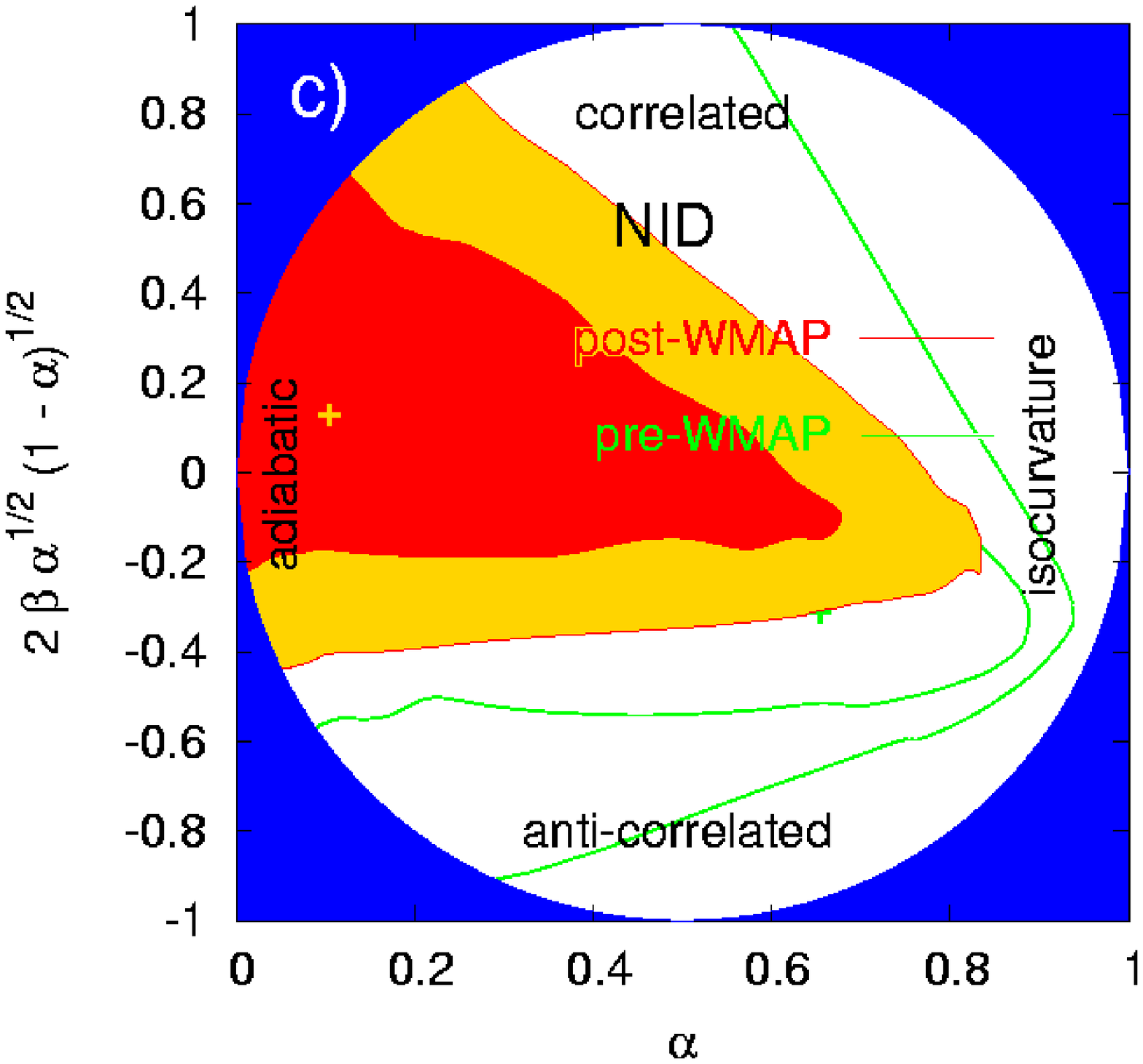,angle=-90,width=7cm}
\psfig{figure=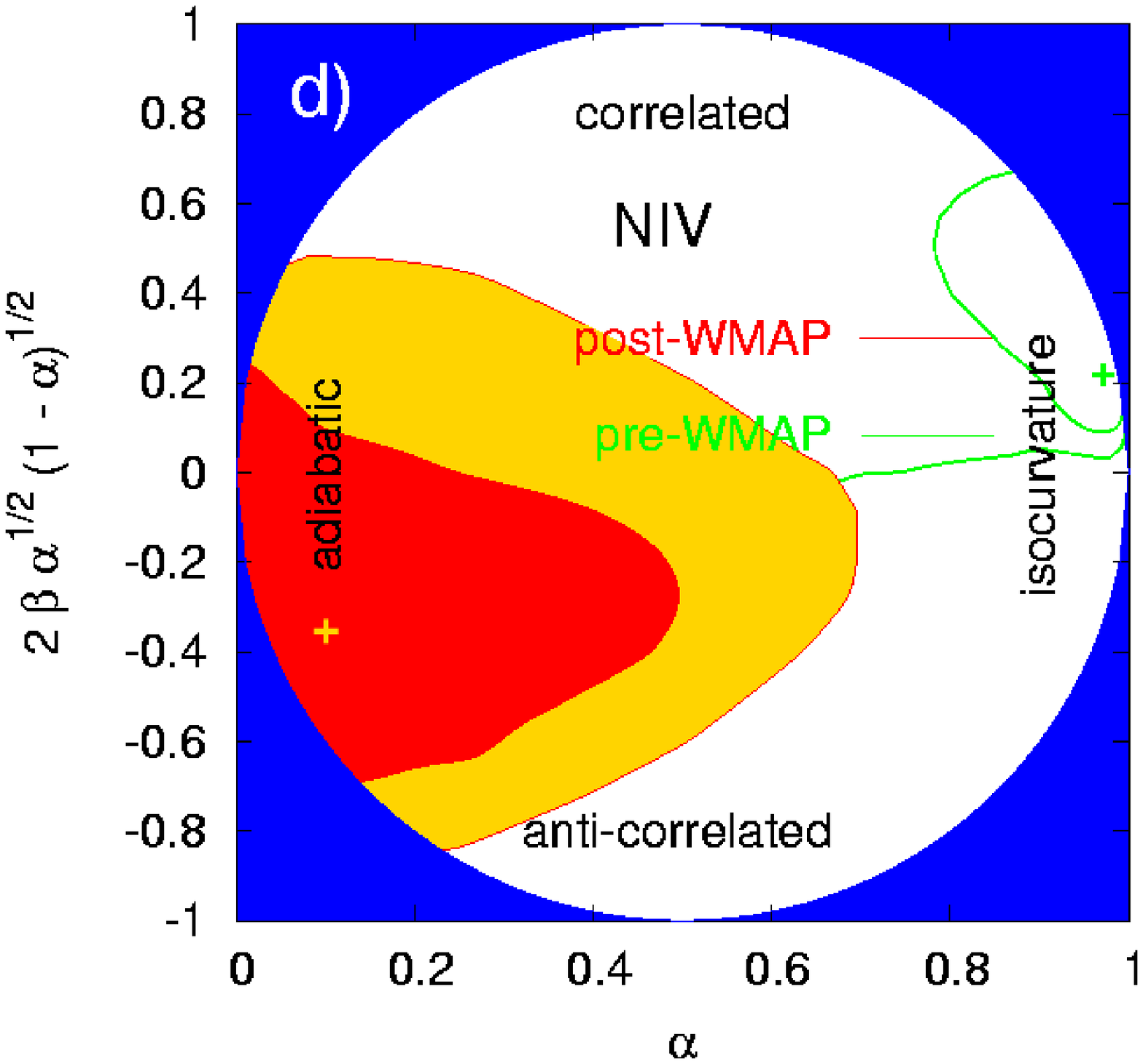,angle=-90,width=7cm}
\vspace{1cm}
\caption{a) the likelihood function of the isocurvature
fraction $\alpha$, for the three different types of uncorrelated
isocurvature modes (i.e., with the prior $\beta = 0$); b) the 1- and
2-$\sigma$ contours of $\alpha$ and the cross-correlated mode
coefficient $2\beta\sqrt{\alpha(1-\alpha)}$, for the CDM
isocurvature mode: the small (red) contours are based on all the data,
with one flat prior $\niso > 0.6$,
while the large (green) ones show the situation before WMAP,
with an additional prior $\omega_B < 0.037$; c) same
as b) for NID; d) same as b) for NIV.
\label{fig:alpha}}
\end{figure}

The CMB temperature fluctuations on large scales, the Sachs-Wolfe
contribution, can be written as
\begin{equation}
{\Delta T\over T} = {1\over3} \Phi^{\rm ad}_{\rm rad} + 
2\Phi^{\rm iso}_{\rm rad} = {1\over5}\Rrad-{2\over5}\Srad\,,
\end{equation}
in terms of the initial conditions during the radiation era,
$\Phi^{\rm ad}_{\rm rad}=(3/5)\Rrad$ and $\Phi^{\rm iso}_{\rm rad}=
-(1/5)\Srad$. However, in order to evaluate the temperature and
polarization anisotropies seen in the CMB today, one has to calculate
the radiation transfer functions for adiabatic and isocurvature
perturbations, $g_l^{\rm ad}(k),\ g_l^{\rm iso}(k)$ and $g_l^{\rm
corr}(k)$. The angular power spectrum for each individual mode is
therefore
\begin{eqnarray}\nonumber
C_l^{\rm ad}\!&\equiv&\!{1\over25}\,\int {dk\over k}\,g_l^{\rm ad}(k)\,
\left({k\over k_0}\right)^{\nad-1}\,,\\[2mm]
C_l^{\rm iso}\!&\equiv&\!{4\over25}\,\int {dk\over k}\,g_l^{\rm iso}(k)\,
\left({k\over k_0}\right)^{\niso-1}\,,\nonumber\\[2mm]
C_l^{\rm corr}\!&\equiv&\!-{4\over25}\,\int {dk\over k}\,g_l^{\rm corr}(k)\,
\left({k\over k_0}\right)^{(\nad+\niso)/2-1}\,,\nonumber
\end{eqnarray}
and the total angular power spectrum is computed as~\cite{Amendola:2001ni}
\begin{equation}
C_l = C_l^{\rm ad} + B^2\,C_l^{\rm iso} + 
2B\,\cos\Delta\,C_l^{\rm corr}\,,
\end{equation}
where $A\equiv1$ has been marginalized over, and $B$ is now the
entropy to curvature perturbation ratio during radiation era,
$B\equiv\Srad/\Rrad$. We will use here a slightly different notation,
used before by other groups~\cite{Langlois:1999dw,Stompor:1995py},
where $A^2 \equiv (1-\alpha)$ and $B^2\equiv\alpha$,
\begin{equation}\label{alphanotation}
C_l = (1-\alpha)\,C_l^{\rm ad} + \alpha\,C_l^{\rm iso} + 2\beta\,
\sqrt{\alpha(1-\alpha)}\,C_l^{\rm corr}\,.
\end{equation}
Here the parameter $\alpha$ runs from purely adiabatic ($\alpha=0$) to
purely isocurvature ($\alpha=1$), while $\beta$ defines the
correlation coefficient, with $\beta=+1/-1$ corresponding to maximally
correlated/anticorrelated modes. There is an obvious relation between
both parametrizations:
\begin{equation}
\alpha = B^2/(1+B^2)\,, \hspace{1cm} \beta = \cos\Delta\,.
\end{equation}
This notation has the advantage that the full parameter space of
$(\alpha,\ 2\beta\sqrt{\alpha(1-\alpha)})$ is contained within a
circle of radius $1/2$. The North and South rims correspond to fully
correlated ($\beta=+1$) and fully anticorrelated ($\beta=-1$)
perturbations, with the equator corresponding to uncorrelated
perturbations ($\beta=0$).  The East and West correspond to purely
isocurvature and purely adiabatic perturbations, respectively. Any
other point within the circle is an arbitrary admixture of adiabatic
and isocurvature modes.

In order to compute the theoretical prediction for the $C_l$
coefficients of the temperature and polarization power spectra, as
well as the matter spectra $P(k)$, for all four different components,
we have used a CMB code developed by one of us (A.R.) which coincides,
within 1\% errors, with the values provided by CMBFAST for AD and CDI
modes, and also includes the neutrino isocurvature modes as well
as the cross-correlated power spectra. Note that the code defines
the tilt of each power spectrum with respect to a pivot
scale $k_0$ corresponding the present value of the Hubble radius, while
CMBFAST uses $k_0 = 0.05$~Mpc$^{-1}$. Thus, the comparison of our
results with those of WMAP~\cite{WMAP} for the CDI mixed model 
is not straightforward.

\begin{figure}
\psfig{figure=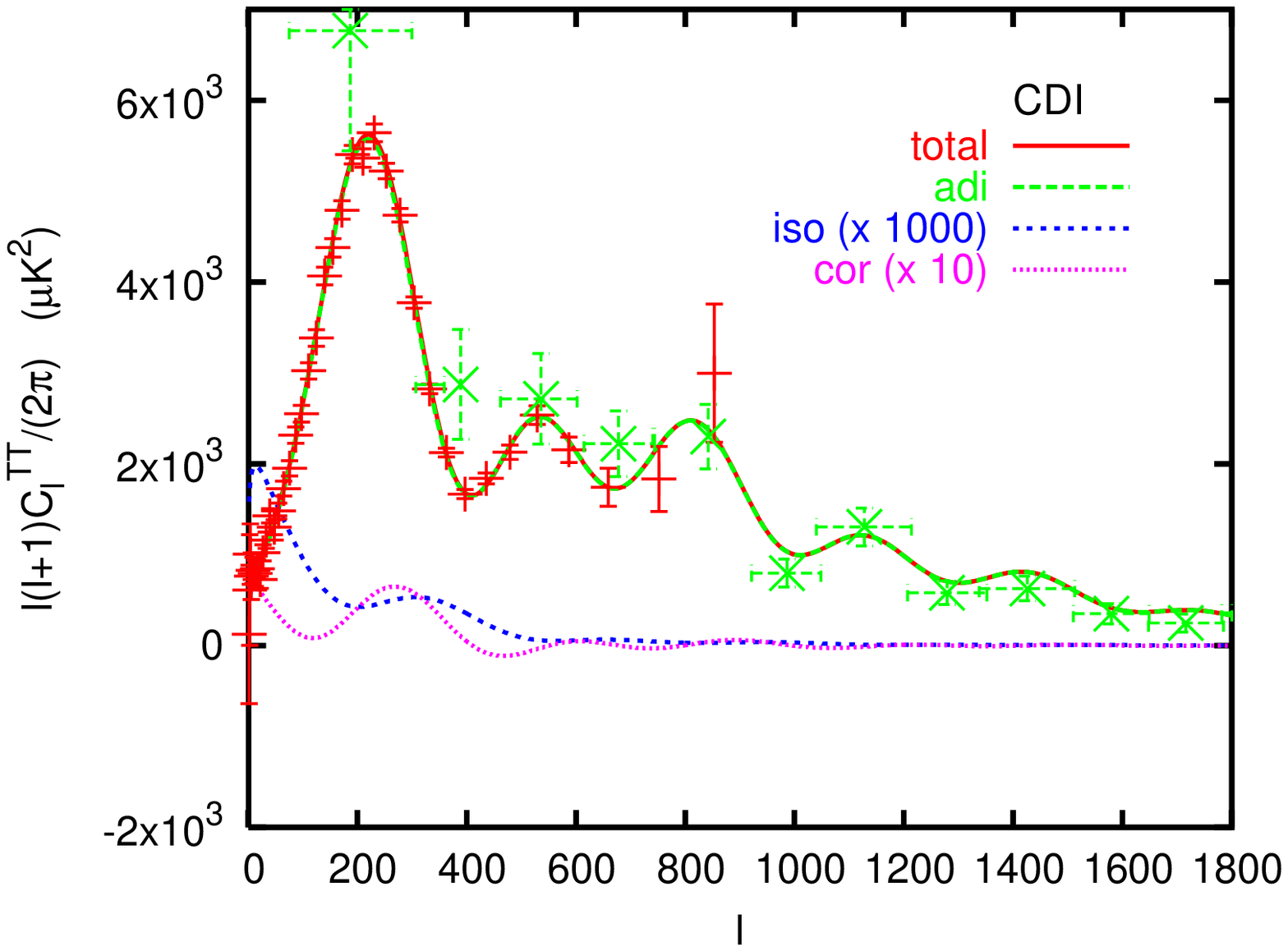,width=8cm}
\psfig{figure=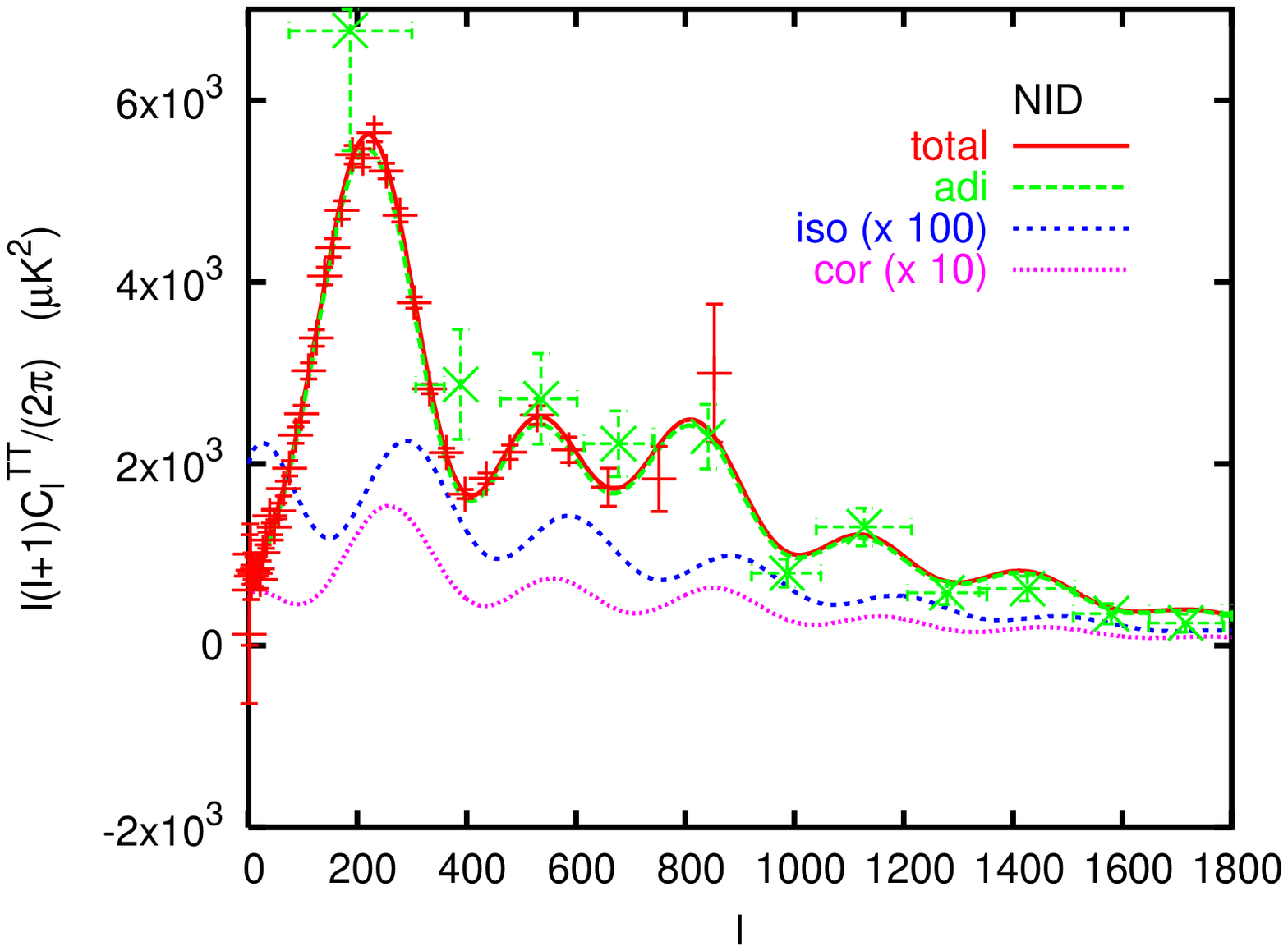,width=8cm}\\
\psfig{figure=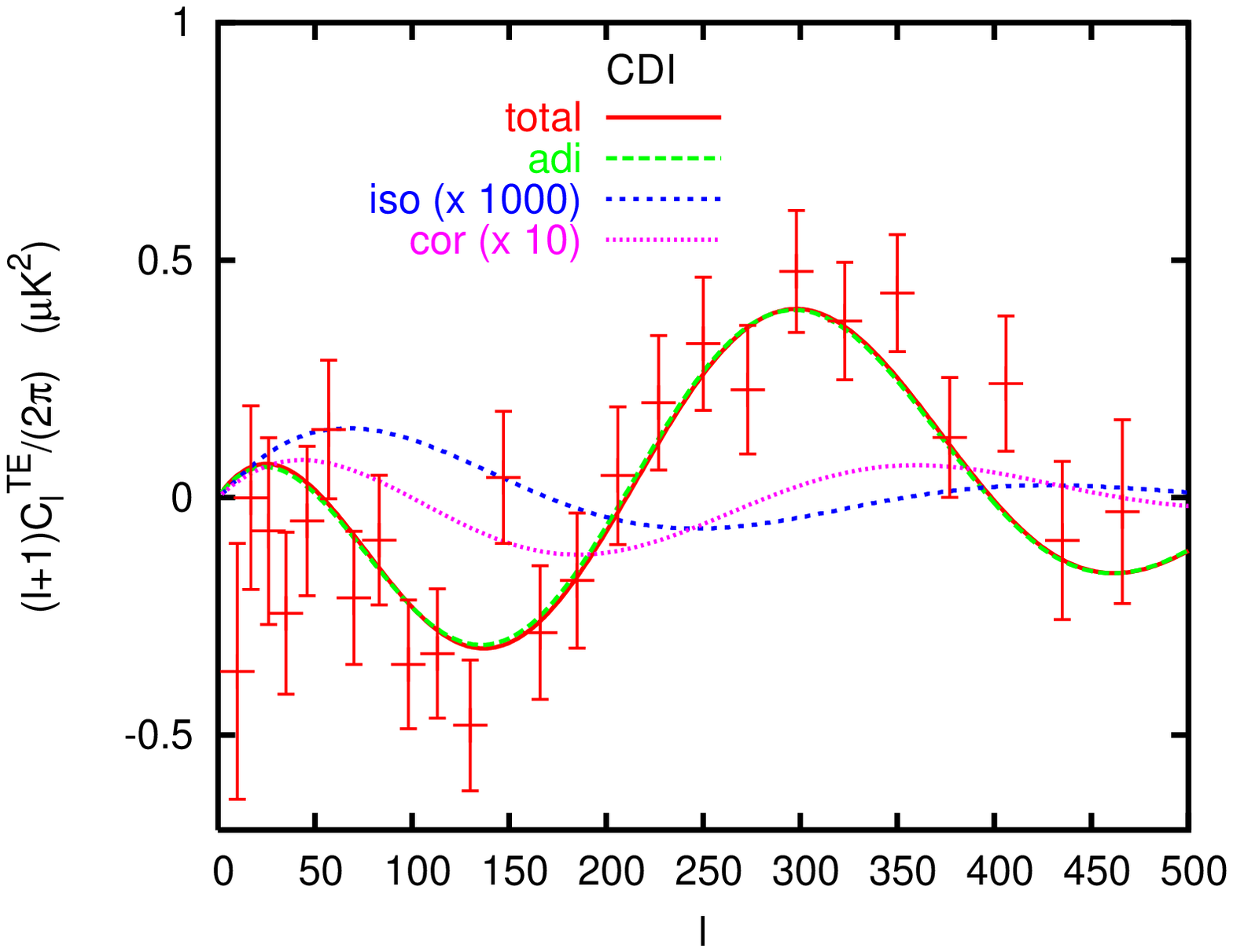,width=8cm}
\psfig{figure=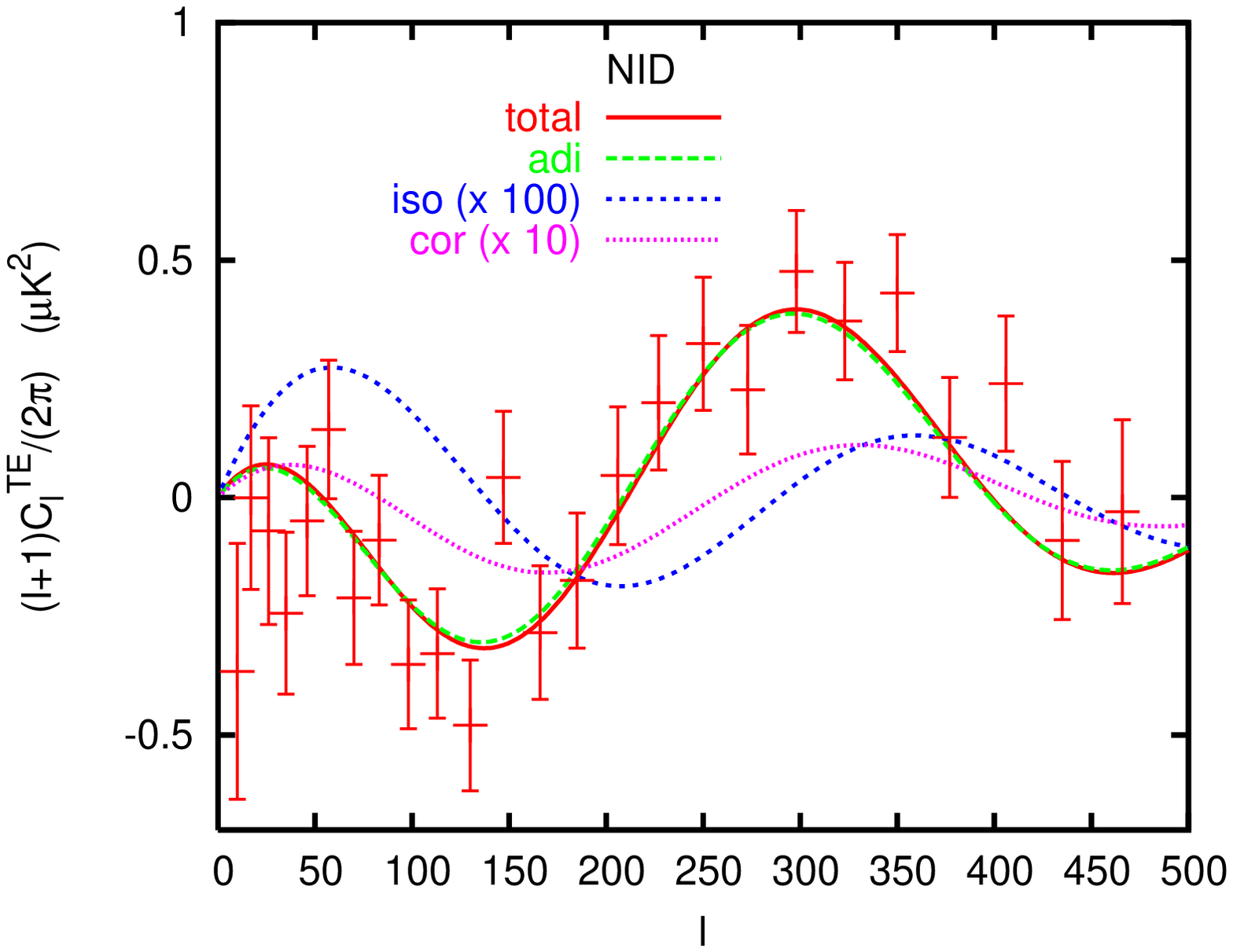,width=8cm}\\
\psfig{figure=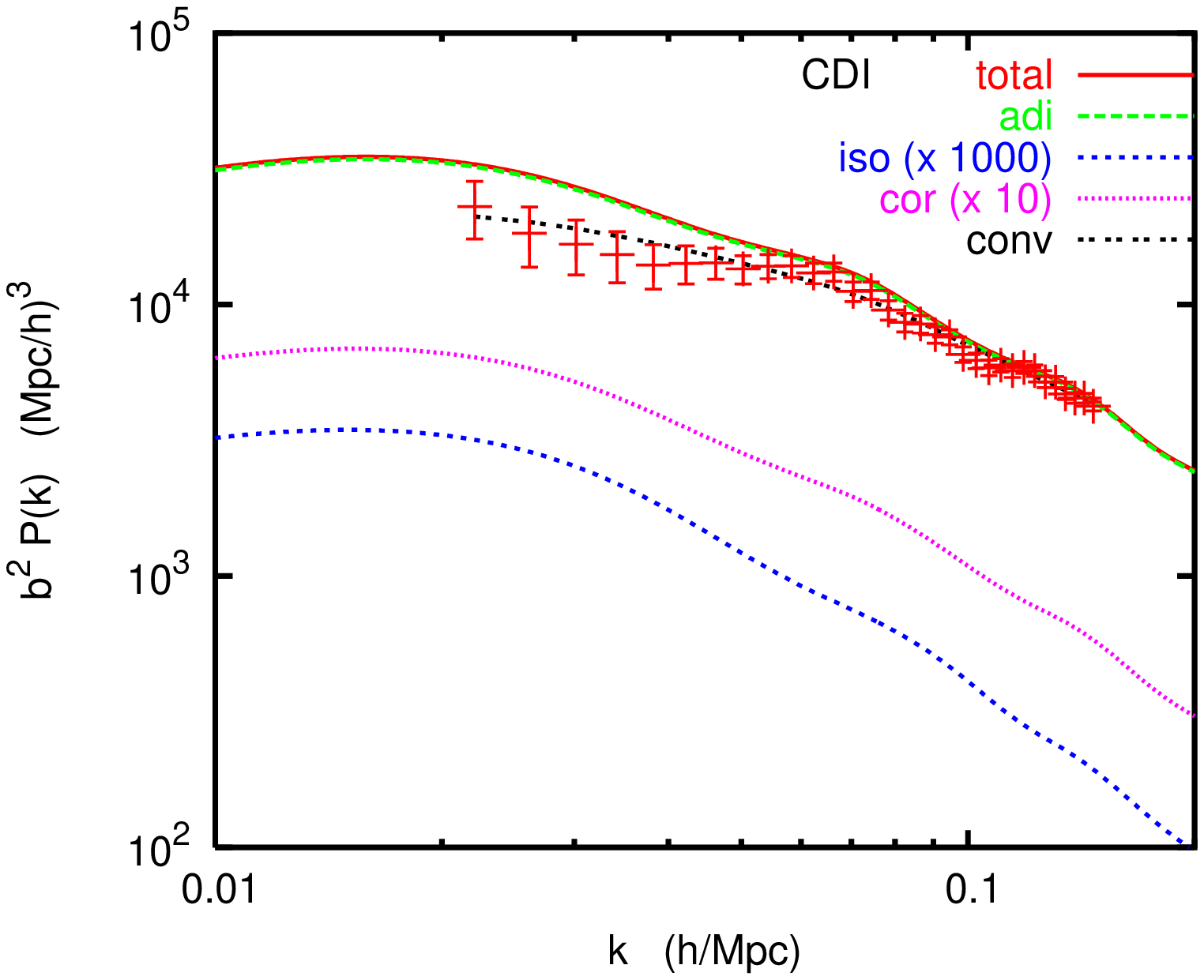,width=8cm}
\psfig{figure=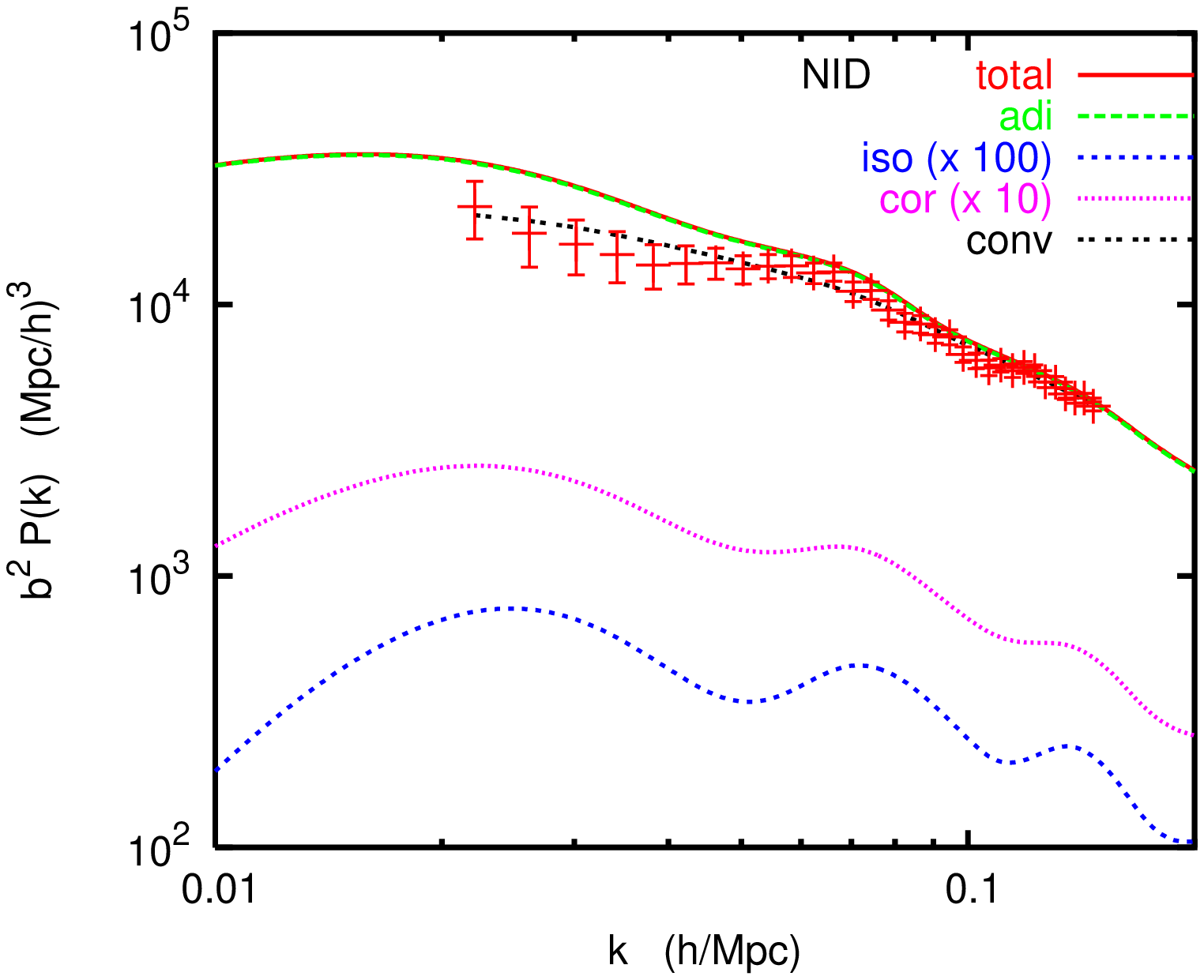,width=8cm}\\
\vspace{1cm}
\caption{ 
Some best fit models compared to data.  The left plots correspond to
correlated CDM isocurvature modes, the right plots to neutrino density
isocurvature modes.  The parameters of these models can be seen in
Table~1. From top to bottom, we show the $\cltt$, $\clte$ and $P(k)$
power spectra, as well as the contribution of each component:
adiabatic (ad), isocurvature (iso), cross-correlated (corr). The last
two components have been rescaled by a factor indicated in each
figure. We also show the data points that we use throughout the
analysis, from WMAP, ACBAR and 2dF. In the case of the matter power
spectrum, we plot in addition the theoretical spectrum convolved with
the experimental window function (conv), in order to allow for a
visual comparison with the data (which is not possible directly with
the unconvolved power spectrum, due to the shape of the window
functions). Note that these two best-fit models are extremely close to
the pure adiabatic one: the differences in the respective power
spectra would be indistinguishable by eye.
\label{fig:bestfit}}
\end{figure}

\begin{table}[t]
\caption{The best fit values of cosmological parameters for both
adiabatic, mixed-isocurvature and correlated models, with their
$\chi^2$ per degree of freedom.\label{bestfit}}
\vspace{0.4cm}
\begin{center}
\begin{tabular}{|l|c|c|c|c|c|c|c|c|}
\hline
model & $\oB$ & $\ocdm$ & $\OL$ & $\nad$ & 
$\niso$ & $\alpha$ & $\beta$ & $\chi^2/\nu$ \\[2mm]
\hline
\hline
AD & 0.021 & 0.12 & 0.70 & 0.95 & $-$ & $-$ & $-$ & 1478.8/1435 \\
\hline
\hline
CDI & 0.023 & 0.12 & 0.73 & 0.99 & 1.02 & 0.10 & $-$ & 1478.3/1433 \\
\hline
BI & 0.023 & 0.12 & 0.73 & 0.99 & 1.02 & 0.72 & $-$ & 1478.3/1433 \\
\hline
NID & 0.023 & 0.12 & 0.73 & 0.99 & 0.95 & 0.37 & $-$ & 1478.2/1433 \\
\hline
NIV & 0.021 & 0.12 & 0.70 & 0.95 & $-$ & 0.0 & $-$ & 1478.8/1433 \\
\hline
\hline
c-CDI & 0.022 & 0.12 & 0.71 & 0.97 & 1.23 & 0.001 & 1.0 & 1478.0/1432 \\
\hline
c-BI & 0.022 & 0.12 & 0.71 & 0.97 & 1.23 & 0.03 & 1.0 & 1478.0/1432 \\
\hline
c-NID & 0.022 & 0.12 & 0.73 & 0.97 & 1.04 & 0.10 & 0.26 & 1477.7/1432 \\
\hline
c-NIV & 0.021 & 0.12 & 0.71 & 0.95 & 0.71 & 0.03 & $-1.0$ & 1477.5/1432 \\
\hline
\end{tabular}
\end{center}
\end{table}

\vspace{1cm}

{\bf Conclusions}. Using the recent measurements of temperature and
polarization anisotropies in the CMB by WMAP, as well as the matter
power spectrum measured by 2dFGRS, one can obtain stringent bounds on
the various possible isocurvature components in the primordial spectrum
of density and velocity fluctuations. We have considered both correlated
and uncorrelated adiabatic and isocurvature modes, and find no
significant improvement in the likelihood of a cosmological model by the
inclusion of an isocurvature component. For instance, a model allowed at
the 2$\sigma$ level by the present observations (WMAP + ACBAR + 2dFGRS)
with a correlated admixture of adiabatic and CDM isocurvature modes,
with $\alpha=0.46$ and $\beta=-0.3$, and tilts $\nad=1.03$ and
$\niso=0.79$, together with cosmological parameters $\OL=0.744$,
$\ocdm=0.112$ and $\oB=0.0246$, has a $\chi^2$ per d.o.f.  of 1.0349,
while the corresponding c-CDI best fit model (see Table.~1) has
$\chi^2/\nu=1.0321$, and the pure adiabatic best fit model has
$\chi^2/\nu=1.0305$. Therefore, we conclude that for the moment there is
no significant improvement in the likelihood of a model by the inclusion
of a small admixture of isocurvature perturbations. In other words, the
basic paradigm of single field inflation passes for the moment all
observational constraints with flying colors. It is
expected~\cite{Trotta:2004sm} that in the near future, with better data, we
will be able to constrain further a possible isocurvature fraction, or
perhaps even discover it.

\section*{Acknowledgments}
This work was supported in part by a CICyT project FPA2000-980, and
by a Spanish-French Collaborative Grant bewteen CICyT and IN2P3.

\end{document}